\documentclass[aps,twocolumn,superscriptaddress]{revtex4-2}
\usepackage[colorlinks=true,bookmarks=true,citecolor=magenta,urlcolor=magenta,linkcolor=magenta,breaklinks]{hyperref} 
\usepackage{breakurl}
\usepackage{sidecap}
\usepackage{amssymb}
\usepackage{hhline}
\usepackage{urwchancal}
\usepackage{xcolor}
\sidecaptionvpos{figure}{t}
\usepackage{amsmath}
\usepackage{graphicx}
\usepackage{esint}
\usepackage{epstopdf}
\usepackage{rotating}
\graphicspath{{pict/}{./}}
\usepackage{bm}%
\usepackage{txfonts}
\usepackage{amsfonts}
\usepackage[classicReIm]{kpfonts}
\usepackage[T1]{fontenc}
\usepackage{microtype,bm,bbm,graphicx,booktabs,times}

\newcounter{Fig}

\newcommand\mymapstol{\mathrel{\ooalign{$\leftarrow$\cr%
  \kern1.75ex\raise0.275ex\hbox{\scalebox{1}[0.4]{$\mid$}}\cr}}}

\newcommand\mymapstor{\mathrel{\ooalign{$\rightarrow$\cr%
  \kern-.15ex\raise.275ex\hbox{\scalebox{1}[0.4]{$\mid$}}\cr}}}
\begin{document}


\title{Magneto-chiral backscatterings by rotationally symmetric nonreciprocal structures}
\author{Chunchao Wen}
\author{Jianfa Zhang}
\email{jfzhang85@nudt.edu.cn}
\author{Shiqiao Qin}
\author{Zhihong Zhu}
\author{Wei Liu}
\email{wei.liu.pku@gmail.com}
\affiliation{College for Advanced Interdisciplinary Studies, National University of Defense Technology, Changsha 410073, P. R. China.}
\affiliation{Nanhu Laser Laboratory and Hunan Provincial Key Laboratory of Novel Nano-Optoelectronic Information Materials and
Devices, National University of Defense Technology, Changsha 410073, P. R. China.}

\begin{abstract}
It was proved that the joint operation of electromagnetic reciprocity and $n$-fold ($n\geq3$) rotational symmetry would secure arbitrary polarization-independent backscattering efficiency [Phys. Rev. B \textbf{103}, 045422 (2021)]. Here we remove the restriction of reciprocity and study the backscatterings of plane waves by rotationally symmetric magneto-optical structures, with collinear incident wavevector, rotational axis and externally applied magnetic field. It is revealed that though nonreciprocity removes the degeneracy of backscattering efficiencies for circularly-polarized incident waves of opposite handedness, the remaining rotational symmetry is sufficient to guarantee that the efficiency is related to the polarization ellipticity only, having nothing to do with the orientations of the polarization ellipses. Moreover, the backscattering efficiency reaches its extremes (maximum or minimum values) always for circularly-polarized incident waves, and for other polarizations the efficiency is their ellipticity-weighted arithmetic average. The principles we have revealed are dictated by rotational symmetries only, which are irrelevant to specific geometric or optical parameters and are intrinsically robust against any rotational-symmetry preserving perturbations. The correlations we have discovered could be further exploited for fundamental explorations in nonreciprocal photonics and practical applications including polarimetry and ellipsometry.
\end{abstract}

\maketitle

\section{Introduction}
\label{section1}

The concept of symmetry has been one of the dominating themes of 20th Century physics, pervading almost all branches of physics and even mathematics~\cite{WEYL_2016__Symmetry,YANG_2013,LEE_1988__Symmetries,RONAN_2007__Symmetry}. In optics and photonics, different sorts of symmetries (discrete and continuous, spatial and temporal, Abelian and non-Abelian)
have been explored for manipulations of light-matter interactions~\cite{BARRON_2009__Molecular,POTTON_Rep.Prog.Phys._reciprocity_2004,FERNANDEZ-CORBATON_2013_Phys.Rev.Lett._Electromagnetica,YANG_2020_ACSPhotonics_Electromagnetic,YAN_2023_Adv.Opt.Photon.AOP_NonAbelian}.
One of the most significant recent developments would be the simultaneous exploitation of parity and time symmetries~\cite{BENDER_1998_Phys.Rev.Lett._Real}, which has effectively reinvigorated the more general field of non-Hermitian physics~\cite{PANCHARATNAM_1955_ProcIndianAcadSci_propagation,BERRY_CURRENTSCIENCE-BANGALORE-_pancharatnam_1994,BERRY_2004_CzechoslovakJournalofPhysics_Physicsa}, leading to tremendous progress in both fundamental investigations and practical applications~\cite{EL-GANAINY_2018_Nat.Phys._NonHermitian,WANG_2023_Adv.Opt.Photon._NonHermitian}. Researches along this trend are rapidly merging with other vibrant fields of topological photonics~\cite{OZAWA_2018_ArXiv180204173,PRICE_2022_J.Phys.Photonics_Roadmap} and singular optics~\cite{NYE_natural_1999,GBUR_2016__Singular,BERRY_2023_LightSciAppl_singularities}, where the core concepts of symmetry, topology and singularity have jointly unlocked enormous unexplored hidden dimensions of freedom~\cite{LIU_ArXiv201204919Phys._Topological,KANG_2023_NatRevPhys_Applications,WANG_2022_Front.Phys._Fundamentals}.  

The study of backscattering and its associated reflection problem has been one of the central topics in optics~\cite{Bohren1983_book,YARIV_2006__Photonics,HORSLEY_Nat.Photonics_spatial_2015,IM_2018_Nat.Photonics_Universal}. A rather outstanding example is the Kerker scattering~\cite{Kerker1983_JOSA} and its generalized form~\cite{LIU_2018_Opt.Express_Generalized}, where the core effect is backscattering suppression or elimination. The original Kerker effect~\cite{Kerker1983_JOSA} can be can be interpreted from the perspective of joint symmetries of rotation and electromagnetic duality~\cite{FERNANDEZ-CORBATON_2013_Opt.ExpressOE_Forwarda,YANG_2020_ACSPhotonics_Electromagnetic}, while its generalized form~\cite{LIU_2018_Opt.Express_Generalized}  from the more generic perspective of phase symmetries (even or odd parities) of electromagnetic multipoles~\cite{Liu2014_ultradirectional,LIU_Phys.Rev.Lett._generalized_2017}. The hidden topological and singular properties that underlie the zero backward scattering of Kerker effect have also been uncovered~\cite{LIU_2017_ACSPhotonics_Beam,CHEN_2019__Singularities}, revealing that the detailed local scattering properties are actually globally constrained by more profound physical and mathematical laws and theorems~\cite{CHEN_2019__Singularities,CHEN_2019_ArXiv190409910Math-PhPhysicsphysics_Linea,CHEN_ACSOmega_Global}.  

It is recently discovered that the sole combination of electromagnetic reciprocity and no less than three-fold rotational symmetry is sufficient to guarantee identical backscattering efficiency for  plane waves of arbitrary polarizations incident along the rotation axis~\cite{CHEN_2021_Phys.Rev.B_Arbitrary}. In this paper, we break the reciprocity by introducing magneto-optical materials and static external magnetic fields. It is revealed that though the nonreciprocity removes the degeneracy of backscattering efficiency for incident left-handed circularly-polarized (LCP) and right-handed circularly-polarized (RCP) waves, the rotation symmetry alone still secures that, besides the strength of the magnetic field applied, the efficiency is decided by the ellipticities of the incident polarization ellipses, having nothing to do with the polarization orientations. Moreover, with a fixed wavelength and external field strength, the efficiency always reaches its maximum or minimum values for circularly-polarized (CP) incident waves, and for other polarizations the efficiency is their weighted arithmetic average (the weight depends on the ellipticity). Similar discussions can be extended from finite scattering structures to extended periodic structures, and the principles we have discovered are naturally applicable to the reflection efficiency. The correlations among incident polarization ellipticities, backscattering (reflection) efficiencies and external magnetic field strengths could be further employed for not only fundamental studies in nonreciprocal photonics but also practical optoelectronic devices for applications including polarimetry and ellipsometry.

\begin{figure}[tp]
\centerline{\includegraphics[width=8.8cm]{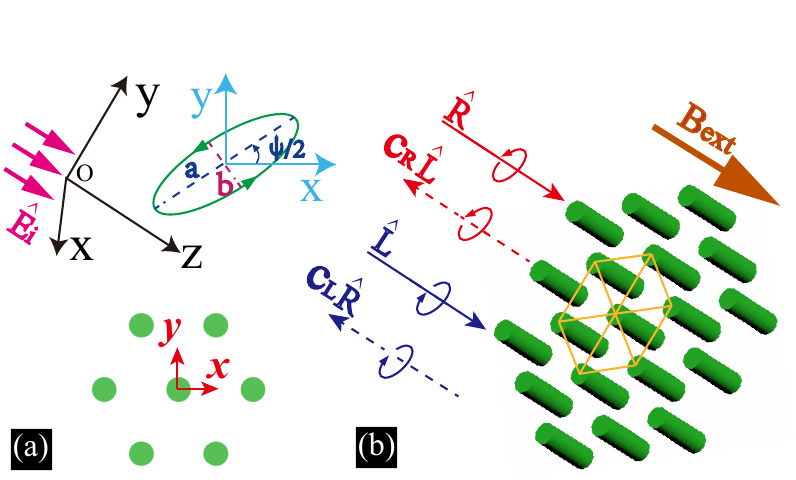}} \caption {\small (a) A plane wave (unit electric field vector $\hat{\mathbf{E}}_i$) incident along $\mathbf{+z}$  direction scattered by rotationally-symmetric (symmetry axis parallel to $\mathbf{z}$-axis) optical structures. For generally incident waves of elliptical polarizations, the polarization ellipses can be characterized by its ellipticy $e=\pm b/a$ (the signs of $\pm$ correspond to polarizations of left and righ handeness, respectively; the inset shows a left-handed polarization ellipse with $e>0$) and  orientation angle with respect to the $\mathbf{x}$-axis is $\psi/2$. (b) Schematic  illustration for the backscattering or reflection with incident CP waves, where the externally applied static magnetic field is parallel to the incident direction and thus also to the rotational symmetry axis. The backscattered or reflected waves are also CP with the helicities flipped.}\label{fig1}
\end{figure} 

\section{Formulism for backscatterings by rotationally symmetric nonreciprocal structures}
\label{section2}

For incident waves of arbitrary polarizations, the normalized electric field unit vector $\hat{\mathbf{E}}_i$ can be expanded into circular basis:

\begin{equation}
\label{arbitrary incident}
\hat{\mathbf{E}}_i=\alpha\hat{\mathbf{L}}+\beta \hat{\mathbf{R}},
\end{equation}
where $\hat{\mathbf{L}}$ and $\hat{\mathbf{R}}$ are respectively LCP and RCP unit vectors; $\alpha$ and $\beta$ are complex expansion coefficients and $|\alpha|^2+|\beta|^2=1$. It is worth mentioning that their relative amplitude $\gamma=|\beta/\alpha|$ decides the ellipticity of the polarization, while their relative phase $\psi$=Arg($\beta/\alpha$) decides the orientation of the polarization ellipses [Fig.~\ref{fig1}(a)]~\cite{YARIV_2006__Photonics,YANG_LPR_Symmetry}. The ellipticity in this work is defined as the ratio between the lengths of the semi-minor and semi-major axes of the polarization ellipses $e=\pm b/a$ [$\pm$ for polarizations of left and right handedness, respectively; see  Fig.~\ref{fig1}(a)] and $\arctan(e)=\pi/4-\arctan(\gamma)$: $e\geq0$ for $\gamma\leq1$ and $e<0$ for $\gamma>1$. For example, the poles and zeros of $|\beta/\alpha|$ correspond to CP singularities where the minor or major axis directions of the polarization ellipses are not well defined ($e=\pm 1$; $\psi$ is not defined); while $|\beta/\alpha|=1$ corresponds to linearly-polarized (LP) singularities where the tangent planes of polarization ellipses are not well defined ($e=0$)~\cite{NYE_natural_1999,GBUR_2016__Singular,BERRY_2023_LightSciAppl_singularities}. For a $n$-fold ($n\geq3$) rotationally-symmetric structure with the incident direction parallel to the rotation axis,  the backscattering processes can be summarized as [also shown schematically in Fig.~\ref{fig1}(b)]:
\begin{equation}
\label{r-coefficient-circular}
\begin{aligned}
\hat{\mathbf{L}} \Rightarrow  c_L \hat{\mathbf{R}};\\
\hat{\mathbf{R}} \Rightarrow c_R \hat{\mathbf{L}},
\end{aligned}
\end{equation}
where $c_L$ and $c_R$ denote backscattering coefficients for incident LCP and RCP waves, respectively. Equation~(\ref{r-coefficient-circular}) basically tells that, upon backscattering CP waves remain to be CP, despite that both their helicities and wavevectors would be flipped (spin angular momentum maintained)~\cite{FERNANDEZ-CORBATON_2013_Opt.ExpressOE_Forwarda,YANG_LPR_Symmetry}.  As a result, combining Eqs.~(\ref{arbitrary incident}) and (\ref{r-coefficient-circular}), the electric field vector of the backscattered wave for an arbitrary incident polarization is:
\begin{equation}
\label{arbitrary-reflection}
\mathbf{E}_{bs}=\alpha c_L \hat{\mathbf{R}}+\beta c_R\hat{\mathbf{L}},
\end{equation}
and the corresponding backscattering efficiency can be expressed as:
\begin{equation}\
\Re_{arb}=|\mathbf{E}_{bs}/\hat{\mathbf{E}}_i|^2=|\alpha c_L|^2 +|\beta c_R|^2.
\end{equation}
If the rotationally symmetric structures are simultaneously reciprocal, it has been proved that reciprocity secures that $c_L =c_R=c$~\cite{CHEN_2021_Phys.Rev.B_Arbitrary} and thus we have
\begin{equation}\
\label{identical-efficiency}
\Re_{arb}=|c|^2 (|\alpha|^2+|\beta|^2)=|c|^2, 
\end{equation}
which means that the backscattering efficiency is the same for incident waves of arbitrary polarizations. For nonreciprocal structures, generally there would be chiroptical backscattering ($\Re_{L}=|c_L|^2 \neq \Re_{R}=|c_R|^2$) and the efficiency for arbitrarily-polarized incident waves  can be reformulated as:
\begin{equation}\
\label{arbitrary-efficiency}
\Re_{arb}=\frac{1}{\gamma^2+1}\Re_L +\frac{\gamma^2}{\gamma^2+1}\Re_R.
\end{equation}
That is, the backscattering efficiency is an ellipticity-weighted arithmetic mean of those for incident LCP and RCP waves $\Re_{arb}\in [\Re_L, \Re_R] $, and it is dependent on the polarization ellipticity and independent on its orientation. For example, for linear polarizations of arbitrary orientations, $\Re_{\rm{LP}}=(\Re_L +\Re_R)/2$.

To further quantify the optical chirality in terms of backscatterings, in a similar fashion to the circular dichroism (CD) factor~\cite{Bohren1983_book}, we have defined: 
\begin{equation}\
\label{cd-bs}
{\rm{CD}}_{bs}=2\frac{\Re_R-\Re_L}{\Re_R+\Re_L},
\end{equation}
which characterizes the relative contrast between backscattering efficiencies for incident CP waves of opposite handedness. We emphasize the conclusions we have drawn so far are independent on the specific geometric or optical parameters of the structures investigated, as long as rotational symmetries required are present.  All the formulism above applies not only to backscatterings by finite scattering bodies, but also to reflections by extended infinite periodic or quasi-periodic structures. 

Throughout this paper, we employ the magneto-optical materials for which the relative permittivity tensor $\varepsilon$ can be expressed as (the external magnetic field $\mathbf{B}_{ext}$ that breaks reciprocity is along \textbf{z}-axis and $\perp$ denotes the components that are perpendicular to $\mathbf{B}_{ext}$ and thus are on the \textbf{x}-\textbf{y} plane):
\begin{equation}
\varepsilon=\left[\begin{array}{ccc}
\varepsilon_{\perp} & -i\varepsilon_{\mathbf{B}} & 0 \\
i\varepsilon_{\mathbf{B}} & \varepsilon_{\perp} & 0 \\
0 & 0 & \varepsilon_{\|}
\end{array}\right].
\end{equation}
For the magneto-optical polar semiconductor material n-doped InSb, the relative dielectric permittivity tensor can be effectively described by the following Drude-like model (neglecting the higher-order terms involving the magnetic field)~\cite{PALIK_1976_Phys.Rev.B_Coupled,MONCADA-VILLA_2015_Phys.Rev.B_Magnetic}:
\begin{equation}
\begin{aligned}
& \varepsilon_{\perp}/\varepsilon_{\infty}=\frac{\omega_L^2-\omega^2-i \Gamma_p \omega}{\omega_T^2-\omega^2-i \Gamma_p \omega}+\frac{\omega_p^2\left(\omega+i \Gamma_f\right)}{\omega\left[\omega_c^2-\left(\omega+i \Gamma_f\right)^2\right]}; \\
& \varepsilon_{\mathbf{B}}/\varepsilon_{\infty}=\frac{\omega_p^2 \omega_c}{\omega\left[\left(\omega+i \Gamma_f\right)^2-\omega_c^2\right]};\\
& \varepsilon_{\|}/\varepsilon_{\infty}=\frac{\omega_L^2-\omega^2-i \Gamma_p \omega}{\omega_T^2-\omega^2-i \Gamma_p \omega}-\frac{\omega_p^2}{\omega\left(\omega+i \Gamma_f\right)}.
\end{aligned}
\end{equation}
Here $\omega$ is the angular frequency of the incident wave; $\varepsilon_{\infty}=15.7$ is the high-frequency dielectric constant; $\omega_L=3.62 \times 10^{13} \mathrm{Hz}$ is the longitudinal optical phonon frequency; $\omega_T=3.39 \times 10^{13} \mathrm{Hz}$ is the transverse optical phonon frequency; $\omega_p=3.14 \times 10^{13}\mathrm{Hz}$ is the plasma frequency;  $\Gamma_p=5.65 \times 10^{11} \mathrm{Hz}$ is the phonon damping constant; $\Gamma_f=3.39 \times 10^{12} \mathrm{Hz}$ is the free carrier damping constant. The breaking of reciprocity is manifest through the non-diagonal asymmetric element $\varepsilon_{\mathbf{B}}$, which is proportional to the cyclotron frequency $\omega_c=e\mathbf{B}_{ext}/m^*$ (here $e$ is the elementary charge and the effective mass is $m^*=0.022~m_0$, where $m_0$ is the electron mass).

\section{Magneto-chiral backscatterings by rotationally symmetric finite scattering structures}
\label{section3}

\begin{figure}[tp]
\centerline{\includegraphics[width=8.8cm]{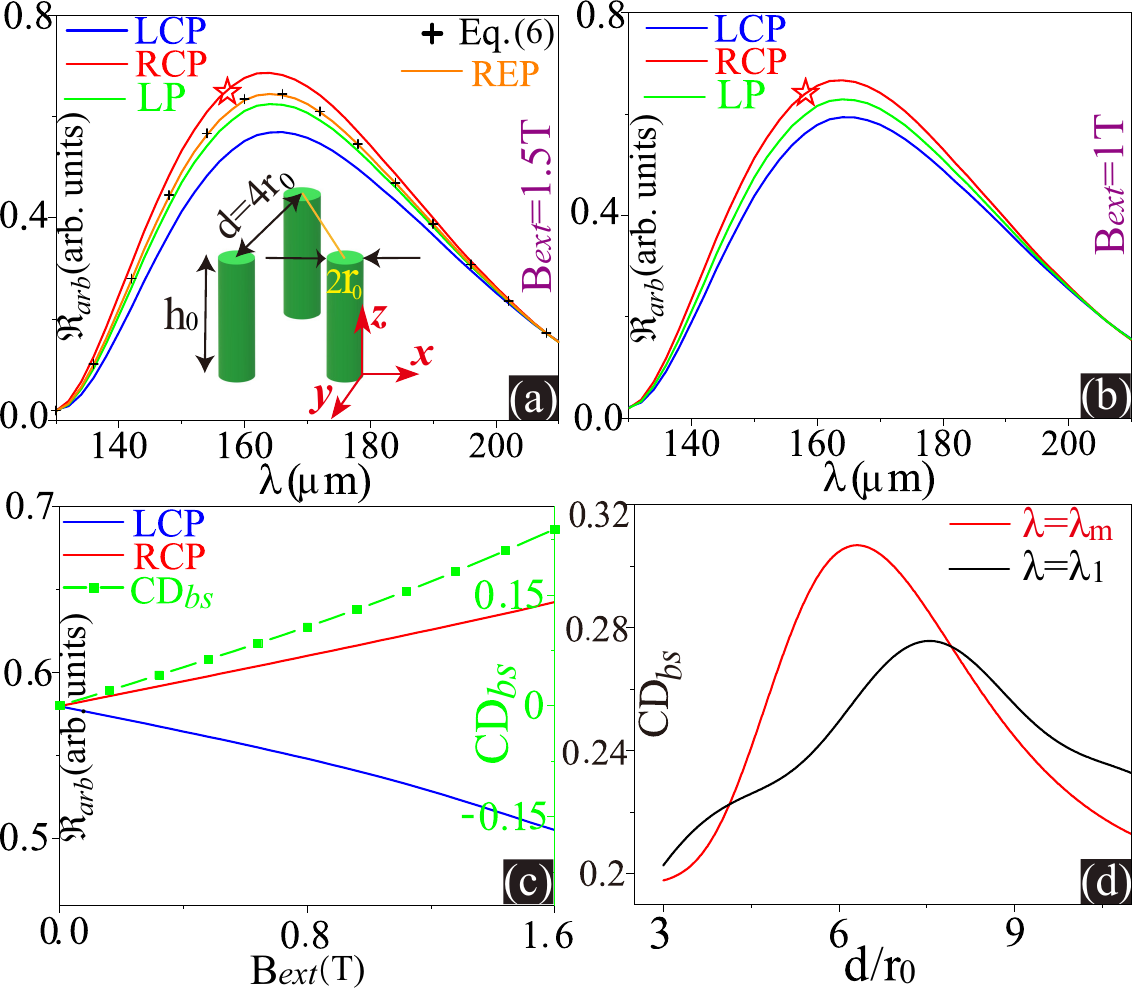}} \caption{\small  (a) Backscattering efficiencies (incident waves are LCP, RCP, LP and REP with $\gamma=\sqrt{2}$) for three magneto-optical circular cylinders centered at the vertices of the regular triangle (edge length $d=4r_0$), with both the cylinder axes and the external static magnetic field ($\mathbf{B}_{ext}=1.5$~T) parallel to the $\mathbf{z}$-axis (see the inset). The star pinpoints the maximum contrast wavelength at the spectral regime investigated. For the REP, two sets of results are shown: numerical (solid orange curve) and those  obtained through Eq.~(\ref{arbitrary-efficiency}) (crosses). (b) Same as (a) except that now the incident waves are LCP, RCP and LP only, with the external magnetic field changed to $\mathbf{B}_{ext}=1$~T. (c) The dependence of backscattering efficiencies (solid red and blue curves for incident RCP and LCP waves, respectively) and ${\rm{CD}}_{bs}$ (green curve) on $\mathbf{B}_{ext}$ with fixed parameters $d=4r_0$ and $\lambda=\lambda_1= 156~\mu$m. (d) The dependence of ${\rm{CD}}_{bs}$ on inter-particle distance $d$ with $\mathbf{B}_{ext}=1.5$~T: black curve for fixed incident wavelength $\lambda=\lambda_1$ and red curve for $\lambda=\lambda_m$ which are different for changing $d$.}
\label{fig2}
\end{figure}

To verify what has been formulated above, we start with the simplest $3$-fold rotationally symmetric configuration: three identical magneto-optical circular cylinders (throughout this paper the geometric parameter of each cylinder is fixed: radius is  $r_0=20~\mu$m and height is $h_0=120~\mu$m) are centered on the vertices of an equilateral triangle (edge length $d=4r_0= 80~\mu$m) with the axes of the cylinders along the $\mathbf{z}$-axis. Such a scattering configuration exhibits also extra mirror symmetries (symmetry planes both parallel and perpendicular to the $\mathbf{z}$-axis), which nevertheless would all be broken by $\mathbf{B}_{ext}$ due to its pseudo-vector nature. In Fig.~\ref{fig2}(a) we show the backscattering efficiencies (numerical results in this work are obtained through commercial software COMSOL Multiphysics) for four incident polarizations [LCP, RCP, LP and a right-handed elliptic polarization (REP) with $\gamma=\sqrt{2}$ and $e\approx-0.17$] with $\mathbf{B}_{ext}$ fixed at $\mathbf{B}_{ext}=1.5~$T. For linear polarizations of arbitrary orientations, as has been formulated $\Re_{\rm{LP}}=(\Re_L +\Re_R)/2$, and this is clear not only here in Fig.~\ref{fig2}(a) [also verified in other structures, as shown in Fig.~\ref{fig2}(b), Figs.~\ref{fig3}(a)-(c) and  Figs.~\ref{fig4}(a)-(c)]. For the REP, the crosses denote results obtained through Eq.~(\ref{arbitrary-efficiency}), which agree perfectly with the numerical results. Another set of results is also shown in Fig.~\ref{fig2}(b) with $\mathbf{B}_{ext}=1$~T, and it is shown the maximum contrast wavelength ($\lambda_m=\lambda_1= 156~\mu$m; it is pinpointed by the star, where the contrast between $\Re_L$  and $\Re_R$ is the largest in the spectral region investigated)  barely shifts with changing $\mathbf{B}_{ext}$.

We then show the dependence of ${\rm{CD}}_{bs}$ on $\mathbf{B}_{ext}$ [obtained at the fixed wavelength ($\lambda_m=\lambda_1\approx 156~\mu$m) for each $\mathbf{B}_{ext}$] in Fig.~\ref{fig2}(c) (the green curve), and it is clear that increasing $\mathbf{B}_{ext}$ would lead to more pronounced chiroptical responses with larger ${\rm{CD}}_{bs}$. Figure~\ref{fig2}(c) also includes the evolutions of backscattering efficiencies with changing $\mathbf{B}_{ext}$, where the solid red and blue curve correspond to incident RCP and LCP waves, respectively. Here we have restricted to $\mathbf{B}_{ext}\geq 0$ ($\mathbf{B}_{ext}$ along the positive $\mathbf{z}$-direction), as parity conservation requires that ${\rm{CD}}_{bs}(\mathbf{B}_{ext})=-{\rm{CD}}_{bs}(-\mathbf{B}_{ext})$. We have also shown in Fig.~\ref{fig2}(d) the dependence of ${\rm{CD}}_{bs}$ on $d$  with $\mathbf{B}_{ext}$ fixed at $\mathbf{B}_{ext}=1.5$~T. In contrast to the scenario of changing $\mathbf{B}_{ext}$, the maximum contrast wavelength $\lambda_m$  is significantly dependent on $d$, and the relation between ${\rm{CD}}_{bs}$ (obtained at  $\lambda_m$) and $d$ are shown in Fig.~\ref{fig2}(d) (the red curve). As is shown, ${\rm{CD}}_{bs}$ would not change monotonically with increasing $d$, reading the peak at some optimum inter-particle distance.  With $d\rightarrow\infty$, ${\rm{CD}}_{bs}$ would asymptotically reach the value for an isolated cylinder. For comparison, in Fig.~\ref{fig2}(d) we also show the evolution of  ${\rm{CD}}_{bs}$ with increasing $d$ at the fixed wavelength of $\lambda=\lambda_1= 156~\mu$m (black curve), which as expected is different from that calculated at $\lambda_m$.

The results for other similar scattering configurations that exhibit $4,5,6$-fold rotational symmetries (identical cylinders are centered on the vertices of squares, regular pentagon and hexagon with the same edge length $d=4r_0$) are summarized in Fig.~\ref{fig3}. Similar to the $3$-fold rotational symmetry scenario: there is no chiroptical backscattering (${\rm{CD}}_{bs}=0$) for $\mathbf{B}_{ext}=0$, as is required by Eq.~(\ref{identical-efficiency}); increasing $\mathbf{B}_{ext}$ would induce more significant nonreciprocity and thus also larger ${\rm{CD}}_{bs}$.

\section{Magneto-chiral reflections by rotationally symmetric infinite periodic structures}

\begin{figure}[tp]
\centerline{\includegraphics[width=8.8cm]{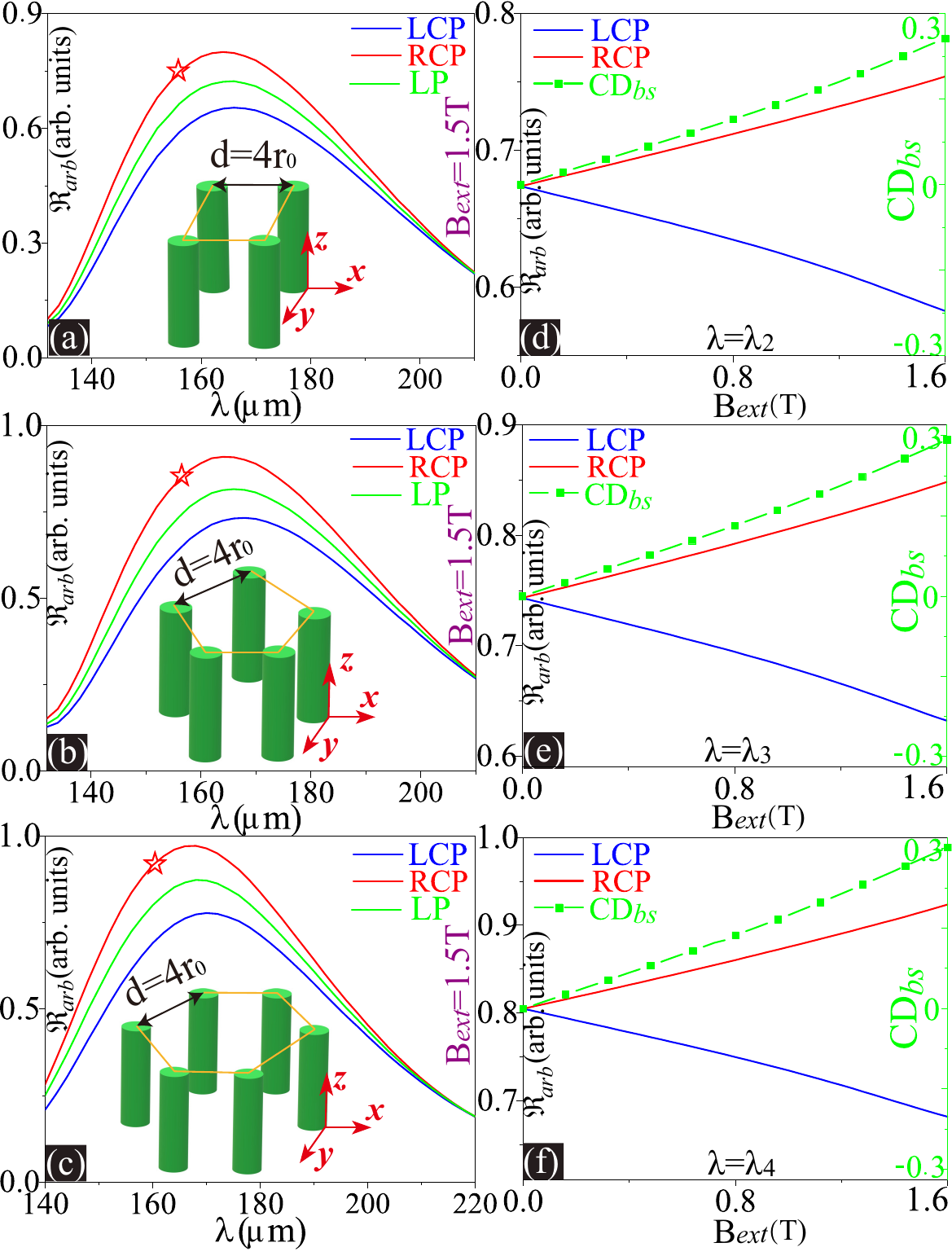}} \caption{\small  (a)-(c) Backscattering efficiencies (incident waves are LCP, RCP, and LP) for  $4-6$-fold rotationally symmetric magneto-optical circular cylinders located respectively at the vertices of square, regular pentagon and hexagon  (edge length $d=4r_0$). All cylinder axes and the external static magnetic field ($\mathbf{B}_{ext}=1.5$~T) are parallel to the $\mathbf{z}$-axis (see insets). The stars pinpoint maximum contrast wavelength $\lambda_m=\lambda_2=156~\mu$m, $\lambda_m=\lambda_3=156~\mu$m and $\lambda_m=\lambda_4=160~\mu$m, for (a)-(c) respectively within the spectral regimes investigated. (d)-(f) The dependence of ${\rm{CD}}_{bs}$ and backscattering efficiencies (calculated at the corresponding $\lambda_m$) on $\mathbf{B}_{ext}$ for the scattering configurations in (a)-(c), respectively. }\label{fig3}
\end{figure}

\begin{figure}[tp]
\centerline{\includegraphics[width=8.8cm]{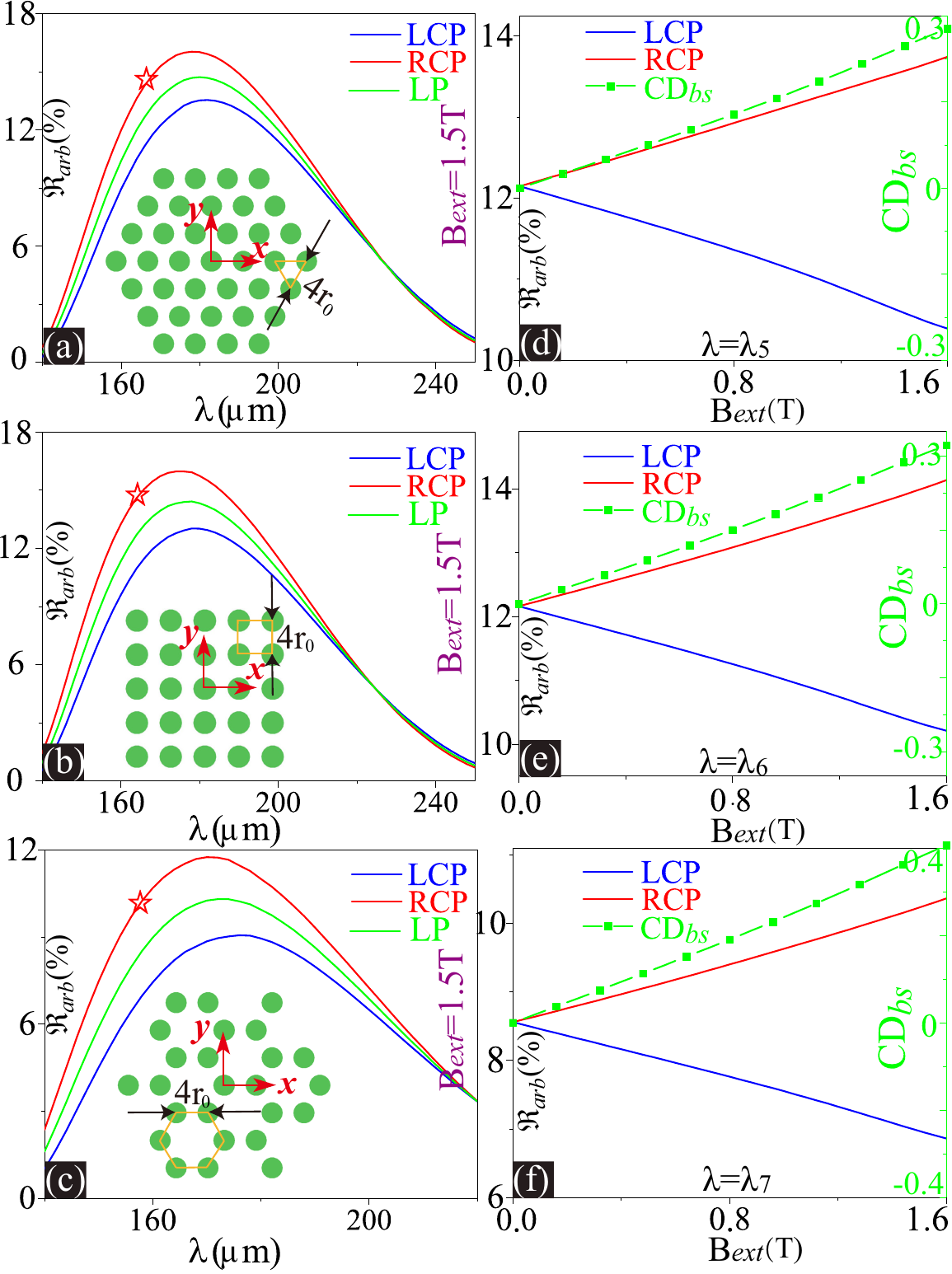}} \caption{\small \small  (a)-(c) Reflection efficiency spectra (incident waves are LCP, RCP, and LP) for  periodic  triangular, square and hexagonal lattices: the unit-cells of those lattices are the finite scattering configurations studied in Fig.~\ref{fig2} (a) and Figs.~\ref{fig3} (a) and (c). All cylinder axes and the external static magnetic field ($\mathbf{B}_{ext}=1.5$T) are parallel to the $\mathbf{z}$-axis (see insets). The stars pinpoint maximum contrast wavelength $\lambda_m=\lambda_5=164~\mu$m, $\lambda_m=\lambda_6=162~\mu$m and $\lambda_m=\lambda_7=158~\mu$m, for (a)-(c) respectively within the spectral regimes investigated. (d)-(f) The dependence of ${\rm{CD}}_{bs}$ and reflection efficiency (calculated at the corresponding $\lambda_m$) on $\mathbf{B}_{ext}$ for the lattices in (a)-(c), respectively.}
\label{fig4}
\end{figure}

We would now extend our discussions from finite scattering structures to extended periodic structures, with the related $3,4,6$-fold rotation symmetries being present. Adopting the scattering structures studied already in the last section [Fig.~\ref{fig2} (a) and Figs.~\ref{fig3} (a) and (c)] as unit cells, we have constructed periodic triangular, square and hexagonal lattices as shown in  Fig.~\ref{fig4}. Our arguments and formulations in Section ~\ref{section3} are based on principles of symmetry only, and thus are also applicable to extended periodic structures. The reflection spectra of those three lattices for three incident polarizations (LCP, RCP and LP) are summarized in Figs.~\ref{fig4}(a)-(c), with $\mathbf{B}_{ext}$ fixed at $\mathbf{B}_{ext}=1.5$~T. As expected, externally applied magnetic field would induced magneto-chiral reflections. For linear polarizations of arbitrary orientations, the relation $\Re_{\rm{LP}}=(\Re_L +\Re_R)/2$ is obviously manifest and for generally elliptically polarized light, the reflection efficiency lie between those for RCP and LCP, as is dictated by Eq.~(\ref{arbitrary-efficiency}). 

For periodic structures, ${\rm{CD}}_{bs}$ can be also defined based on chiral reflection contrast in a similar manner to Eq.~(\ref{cd-bs}). Like finite scattering structures, the ${\rm{CD}}_{bs}$ is calculated at maximum contrast wavelength $\lambda_m$, where the contrast between reflection efficiencies ($|\Re_R-\Re_L|$) is the largest at the spectral regime of interest. For the three lattices discussed above, the dependencies of ${\rm{CD}}_{bs}$ on $\mathbf{B}_{ext}$ are shown in Figs.~\ref{fig4}(d)-(f). As is shown, increasing $\mathbf{B}_{ext}$ would result in larger ${\rm{CD}}_{bs}$, indicating more pronounced nonreciprocity and magneto-chiral responses in terms of reflections.

\section{Conclusion and Perspective}

To conclude, we study the backscattering or reflection properties of magneto-optical structures that exhibit $n$-fold ($n\geq3$) rotation symmetries. It has been previously proved that for waves incident along the rotation symmetry axis, the backscattering or reflection efficiency is fully independent of the polarizations, as secured by the electromagnetic reciprocity.  In this study, we investigate magneto-optical structures and introduce external static magnetic fields to break the reciprocity. It is revealed that when the magnetic field is parallel to the rotation axis (without breaking the geometric rotation symmetry), the backscattering or reflection efficiency is related to the ellipticities of the polarization ellipses only, fully irrelevant to their orientations. Moreover, for generally elliptic polarizations, backscattering or reflection efficiencies are the weighted arithmetic averages of the efficiencies for CP incident waves, with the weight decided by the ellipticity of the polarization ellipse. That is to say, the backscattering or reflection efficiencies always reach their maximum or minimum vales for incident waves of circular polarizations. Our conclusions are based on principles of symmetry only and thus their validity does not rely on specific geometric or optical parameters of the structures as long as the rotation symmetry is preserved. Similar discussions can be extended to waves of other forms for which there is an extra (synthetic) polarization degree of freedom to exploit. 

What about scatterings along the other direction of rotation axis, that is, the forward scattering?  It is previously revealed that for CP incident waves, the forward scatterings are also CP of the same handedness~\cite{FERNANDEZ-CORBATON_2013_Opt.ExpressOE_Forwarda,YANG_LPR_Symmetry}. In contrast to the backscatterings, rotational symmetry and reciprocity are not sufficient to guarantee arbitrary polarization-independent forward scattering efficiency. That is, with the presence of rotational symmetry and reciprocity,  geometric chirality of the structure can be manifested through the  forward scatterings which are incident polarization-dependent~\cite{YANG_LPR_Symmetry}; while the backward scatterings are fully immune to such chirality.  It would be interesting to investigate the competition between geometric chiroptical responses and magneto-chiroptical responses mediated by  rotation symmetries. Furthermore, the electromagnetic duality can be also introduced to render another degree of flexibility~\cite{ FERNANDEZ-CORBATON_2013_Phys.Rev.Lett._Electromagnetica,YANG_2020_ACSPhotonics_Electromagnetic}, where the interplay among symmetries and/or asymmetries of geometry, duality, and reciprocity would provide a broader platform on which new directions for light-matter interaction manipulations can be explored.\\

\section*{acknowledgement}
This research was funded by the National Natural Science Foundation of
China (12274462, 11674396, and 11874426) and the Science and Technology
Planning Project of Hunan Province (2018JJ1033 and 2017RS3039). 



\bibliographystyle{osajnl}
\bibliography{References_scattering5}

\end{document}